\documentclass[tightenlines,superscriptaddress,twocolumn,prl,notitlepage,nofootinbib,preprintnumbers]{revtex4-1}

\usepackage{graphicx}
\usepackage{amsmath}
\usepackage{amsfonts}
\usepackage{amssymb}
\usepackage{float}
\usepackage{hyperref}
\usepackage{multirow}
\usepackage{bbold}
\usepackage{color}
\usepackage{soul}
\usepackage{makecell}
\usepackage{minibox}
\usepackage{soul}
\usepackage{fancyhdr}
\usepackage{enumitem}

\newcommand{\CGMF}{\mathtt{CGMF}}
\newcommand{\TKE}{\langle\mathrm{TKE}\rangle}

\begin{document}

\preprint{LA-UR-17-27790}
\title{Identifying Inconsistencies in Fission Product Yield Evaluations with Prompt Neutron Emission}

\author{Patrick Jaffke}
\email{corresponding author: pjaffke@lanl.gov}
\affiliation{Los Alamos National Laboratory, Los Alamos, NM 87545, USA}

\date{\today}

\begin{abstract}
We present a self-consistency analysis of fission product yield evaluations.
Anomalous yields are determined using a series of simple conservation checks
and comparing charge distributions with common parameterizations. The total
prompt neutron multiplicity as a function of product mass $\bar{\nu}_T(A)$ is
derived directly from the independent fission product yields using average
charge conservation. This method is checked against Monte Carlo simulations of
the de-excitation of the fission fragments in a Hauser-Feshbach statistical
decay framework. The derived $\bar{\nu}_T(A)$ is compared with experimental data, when
available, and used to compare the prompt neutron multiplicity $\bar{\nu}$ for
the various evaluations. Differences in $\bar{\nu}$ for each evaluation are
investigated and possible sources are identified. We also identify fission
reactions that are inconsistent with prompt neutron data and propose possible
solutions to remedy the observed inconsistencies.
\end{abstract}

\maketitle


\section{\label{sec:Intro}Introduction}

Fission product yields are utilized in a wide scope of physics research
and applications. Evaluations of the independent (after prompt neutron
emission but prior to any $\beta$-decays) and cumulative (after prompt
neutron emission and after all $\beta$-decays) fission yields exist for
many fission reactions and a few incident neutron energies in three major
libraries: ENDF-B/VII.1~\cite{chadwick2011endf}, JENDL-4.0u2~\cite{shibata2011jendl},
and JEFF-3.1.1~\cite{koning2006jeff}. These yields are used in depletion
calculations in reactor simulations~\cite{SCALE}, for decay heat
evaluations~\cite{algora2010reactor}, criticality studies~\cite{kodeli2009evaluation},
and reactor design~\cite{rimpault2012needs}. In addition, the fission yield
libraries are used in antineutrino summation calculations~\cite{Fallot:2012jv},
fission theory~\cite{Talou:2014qba,vogt2012event}, and even astrophysics
calculations of the r-process~\cite{Kajino:2016pia}. With such a broad scope
of physical applications, the evaluation libraries undergo scrutinous
self-consistency checks. Even so, inconsistencies can arise. For example,
the decay data involving the $\beta$-decay branching ratios and the
$\beta$-delayed neutron emission probabilities connect the independent and
cumulative fission yields~\cite{kawano2013estimation}. Thus, when new decay
data were incorporated into the ENDF-B/VII.1 sublibrary, but legacy fission yields
were kept, an inconsistency developed between the independent and cumulative
yields~\cite{pigni2015investigation}. Here, we present a consistency check
between the independent fission yields (IFY) and prompt neutron emission.

Many measurements of the prompt neutron emission and fission yields are
conducted independently from one another due to the different detector
requirements for each study. However, correlated measurements of fragment
yields and prompt neutrons have been employed in the past~\cite{oed1984mass}
and in some more recent $2$v\,--\,$2$E measurements~\cite{Meierbachtol201559,Fregeau201635}
or time-of-flight measurements, typically with specially designed ionization
chambers, in conjunction with high-efficiency neutron detectors~\cite{Gook2016366}.
While radiochemical measurements~\cite{prakash1990radiochemical} can
provide a higher degree of accuracy for the yields, they only measure cumulative
fission yields (CFY) and, thus, are more difficult to use for self-consistency checks
between prompt neutron emission and product yields. As many of the fission yields
evaluations were conducted prior to the use of these correlated measurements,
one must ask if they are consistent with prompt neutron data. We aim to answer
this question in the following analysis.

\section{\label{sec:Basic}Basic Yield Checks}
Basic conservation checks allow us to determine the status of the various yield
libraries. For example, we know that the strict mass conservation rule requires that
$\bar{\nu} = A_0 - \sum_A [A\times Y(A)]$, where $A_0$ is the nuclear mass of the
parent fissioning nucleus, $Y(A)$ are the IFY for a product with mass $A$, and
$\bar{\nu}$ is the average prompt neutron multiplicity. After applying this quick
check to all fission reactions and incident neutron energies supported by the three
evaluation libraries, we see that many evaluations are not consistent with prompt
neutron data, but the corresponding evaluation errors are usually larger than one
neutron per fission. This has led us to search for a more precise method of determining the
agreement between fission yields and prompt neutron data.

\begin{figure}[h]
\centering
\includegraphics[width=\columnwidth]{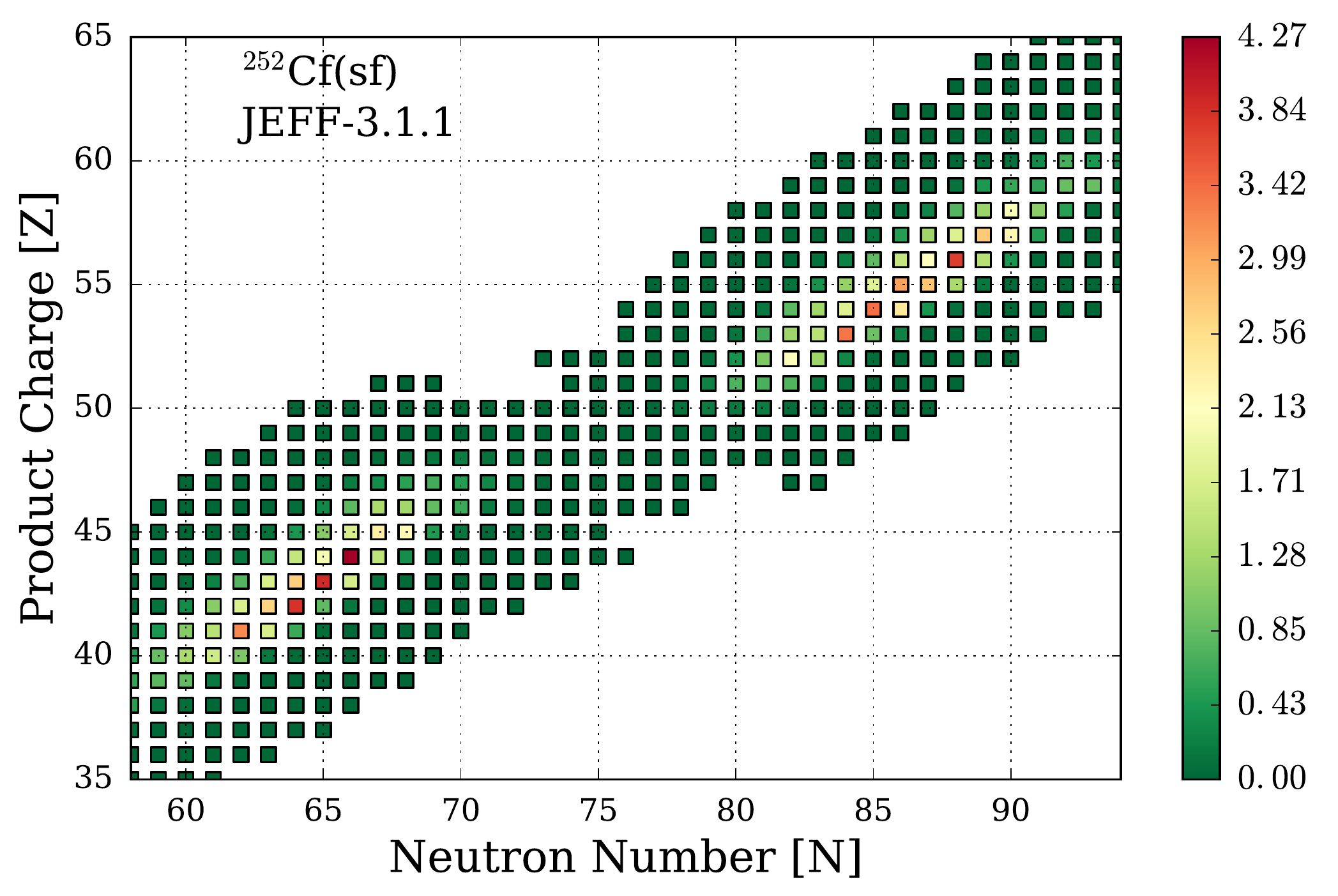}
\caption{\label{fig:Gimpact} Gaussian impact factor for $^{252}$Cf(sf) from
  the JEFF-3.1.1 evaluation. Large impact values denote that a nuclide
  yield deviates from its Gaussian fit and comprises a sizable yield.
  We observe several nuclides which satisfy $G(A,Z)>1$.
  Other fission reactions and evaluations showed fewer nuclides
  with large $G(A,Z)$ values.
 }
\end{figure}

First, we analyze the charge distributions to identify specific nuclides
with anomalous yields. The mass-dependent charge distributions are approximately
Gaussian in product charge $Z$ following Wahl~\cite{wahl2002systematics}
\begin{equation}
\mathcal{I}_A(Z) = \frac{N\times F_{Z,A}}{\sqrt{2\pi\sigma^2}}\exp{-[Z- Z_p(A)]^2/[2\sigma^2]},
\label{eq:Zdistr}
\end{equation}
where the even-odd shell effects are encoded in the term $F_{Z,A}$ (see
Ref.~\cite{wahl2002systematics} for details) and the normalization coefficient $N$
ensures that the integral of $\mathcal{I}_A(Z)$ over $Z$ is unity. Using the various
libraries, we fit Eq.~\ref{eq:Zdistr} to the evaluated charge distributions,
determining best-fit values for $Z_p$ and $\sigma$. One can identify the yields
that do not follow this Gaussian trend by using the impact parameter
\begin{equation}
G(A,Z) = \frac{I^\prime(A,Z)}{2} \mathrm{max} \bigg( \frac{I^\prime(A,Z)}{\mathcal{I}_A(Z)},\frac{\mathcal{I}_A(Z)}{I^\prime(A,Z)} \bigg),
\label{eq:GImpact}
\end{equation}
which represents the ratio of the evaluated fractional yield $I^\prime(A,Z)$ for a given product mass
and charge to the theoretical parameterization. We multiply this ratio by the
fractional yield, which highlights the nuclides that have large yields and are far from
the Gaussian trend. From Eq.~\ref{eq:GImpact}, one can see that a nuclide that
deviates from its expected value of Eq.~\ref{eq:Zdistr} by a factor larger than its
fractional yield generates a $G(A,Z)>1$. A visual representation of this is given by
Fig.~\ref{fig:Gimpact}, where we display $G(A,Z)$ for the JEFF-3.1.1 evaluation of
$^{252}$Cf(sf).
The JEFF-3.1.1 evaluation for $^{252}$Cf(sf) represents a bit of
an anomaly, with $8$ nuclides having $G(A,Z)>3$. Several nuclides appeared
multiple times and are noted in the conclusion. Some nuclides, such as $^{86}$Ge,
have been identified as problematic in other studies~\cite{sonzogni2016effects}.

\begin{figure}[h]
\centering
\includegraphics[width=\columnwidth]{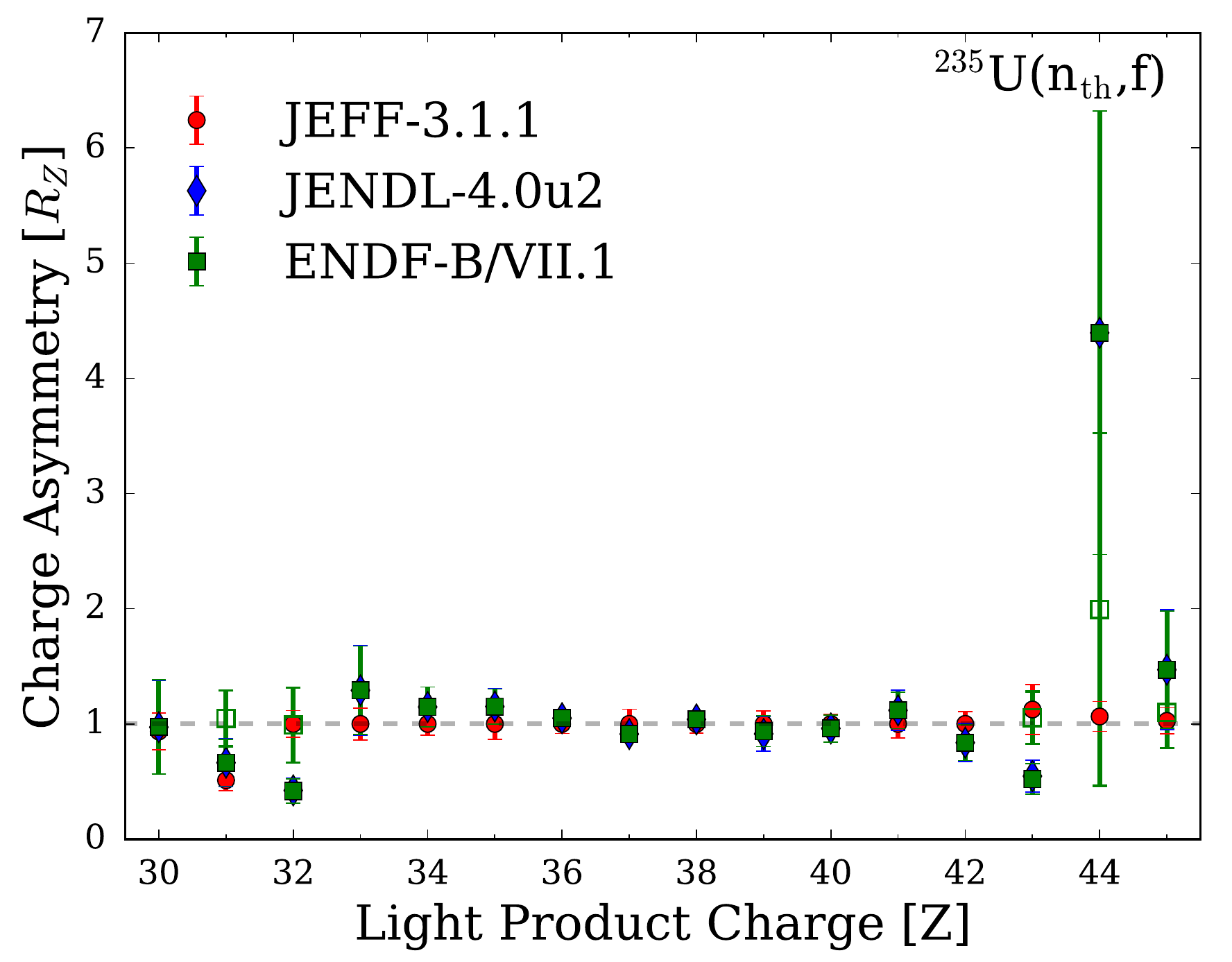}
\caption{\label{fig:U235zasymm} Charge asymmetry for $^{235}$U(n$_\mathrm{th}$,f).
  We compare the three major evaluations~\cite{chadwick2011endf,shibata2011jendl,koning2006jeff}.
  Perfect symmetry corresponds to the line at unity and the open squares are
  the result of manually adjusting a select few yields in ENDF-B/VII.1 to follow the
  Gaussian charge distribution shown in Eq.~\ref{eq:Zdistr}.
 }
\end{figure}

The charge asymmetry $R_Z$ is defined as the ratio of the sum of independent yields
$I(A,Z)$ for a
product charge and its charge complement defined as $Z_c = Z_0 - Z$, where
$Z_0$ is the charge of the fissioning nucleus:
\begin{equation}
R_Z = \displaystyle\sum_A I(A,Z_c) \bigg/ \displaystyle\sum_A I(A,Z).
\label{eq:chargeasymm}
\end{equation}
For fission reactions with no prompt charged particle emission,
Eq.~\ref{eq:chargeasymm} should be unity for all $Z$. We have calculated the
charge asymmetry for all fission reactions and across the three libraries. An
example is given as Fig.~\ref{fig:U235zasymm} for $^{235}$U(n$_\mathrm{th}$,f).
We determine that the majority of isotopes show excellent agreement with unity.
Any deviation is typically in the far-wings or symmetric region, where the yields,
and impact, are lower. We investigated the particular case of
$^{235}$U(n$_\mathrm{th}$,f) in the ENDF-B/VII.1 library. Upon inspection of
the charge distributions, the yields for some cadmium, promethium, and indium
isotopes were far above the Gaussian parameterization. Lowering
these yields to follow Eq.~\ref{eq:Zdistr} results in the open squares in
Fig.~\ref{fig:U235zasymm}; now consistent with unity.

\section{\label{sec:NuDerive}Prompt Neutron Derivation}

As mentioned before, one can trivially determine the average prompt neutron multiplicity
$\bar{\nu}$ by comparing the parent nucleus mass and the weighted mass yields, but
the errors are large enough to allow consistency with all experimental values. A more
precise derivation comes from using the less-stringent rule of charge conservation on
average, given by the following procedure:
\begin{enumerate}
\item Fit the charge distributions of the IFY with Eq.~\ref{eq:Zdistr}.
\item Use the best-fit parameters to determine the average charge $\langle Z\rangle$
	as a function of product mass $A$.
\item Apply average charge conservation to infer mass pairs of $A$ and its complement
         $A_c$.
\item Determine the total average prompt neutron multiplicity $\bar{\nu}_T$ between $A$ and $A_c$.
\end{enumerate}
Without the even-odd shell effects, $\langle Z\rangle$ would be precisely equal to
$Z_p$ in Eq.~\ref{eq:Zdistr}. The parameterization of the even-odd effects has a
negligible effect, but we have included them from
Wahl~\cite{wahl1962nuclear,wahl2002systematics} for completeness.

Charge conservation on average is represented by $\langle Z\rangle (A) + \langle Z\rangle(A_c) = Z_0$.
This requirement creates
mass pairs $A$ and $A_c$, which we use to solve for the average total prompt
neutrons emitted $\bar{\nu}_T(A) = A_0 - A - A_c$ between a product mass $A$ and
its complement. We note that this procedure is well-defined when there is a single
$A_c$ satisfying the charge conservation. This occurs because the $\langle Z\rangle(A)$
functions are monotonically increasing, but there are a few instances where this is
not satisfied. Only some evaluations exhibited this problem and it always occurred
for one or two masses in the symmetric region. The errors of these were included in our
analysis. The total neutrons emitted between the two products is
\begin{equation}
\bar{\nu}_T(A) = \bar{\nu}(A) + \bar{\nu}(A_c).
\label{eq:nubarT}
\end{equation}
Then, one can use the $\bar{\nu}_T(A)$ and the IFY to compute the average prompt
neutron multiplicity per fission
\begin{equation}
\bar{\nu} = \displaystyle\sum_{A} \bar{\nu}_T(A) \times Y(A),
\label{eq:Nubar}
\end{equation}
where $A$ runs over the light or heavy masses.

We use the $\CGMF$ code~\cite{Becker:2013vk}, a Monte Carlo
implementation of the Hauser-Feshbach statistical theory to calculate the de-excitation of
pre-neutron emission fission fragments. This implementation has been shown to reproduce
many fission observables in very reasonable agreement with experimental data~\cite{talou2011advanced}.
The benefit of utilizing this model is that the prompt neutron observables and fission product
yields are generated self-consistently. Thus, a perfect reconstruction of $\bar{\nu}_T(A)$
from the simulated IFY should match the simulated one.

\begin{figure}[h]
\centering
\includegraphics[width=\columnwidth]{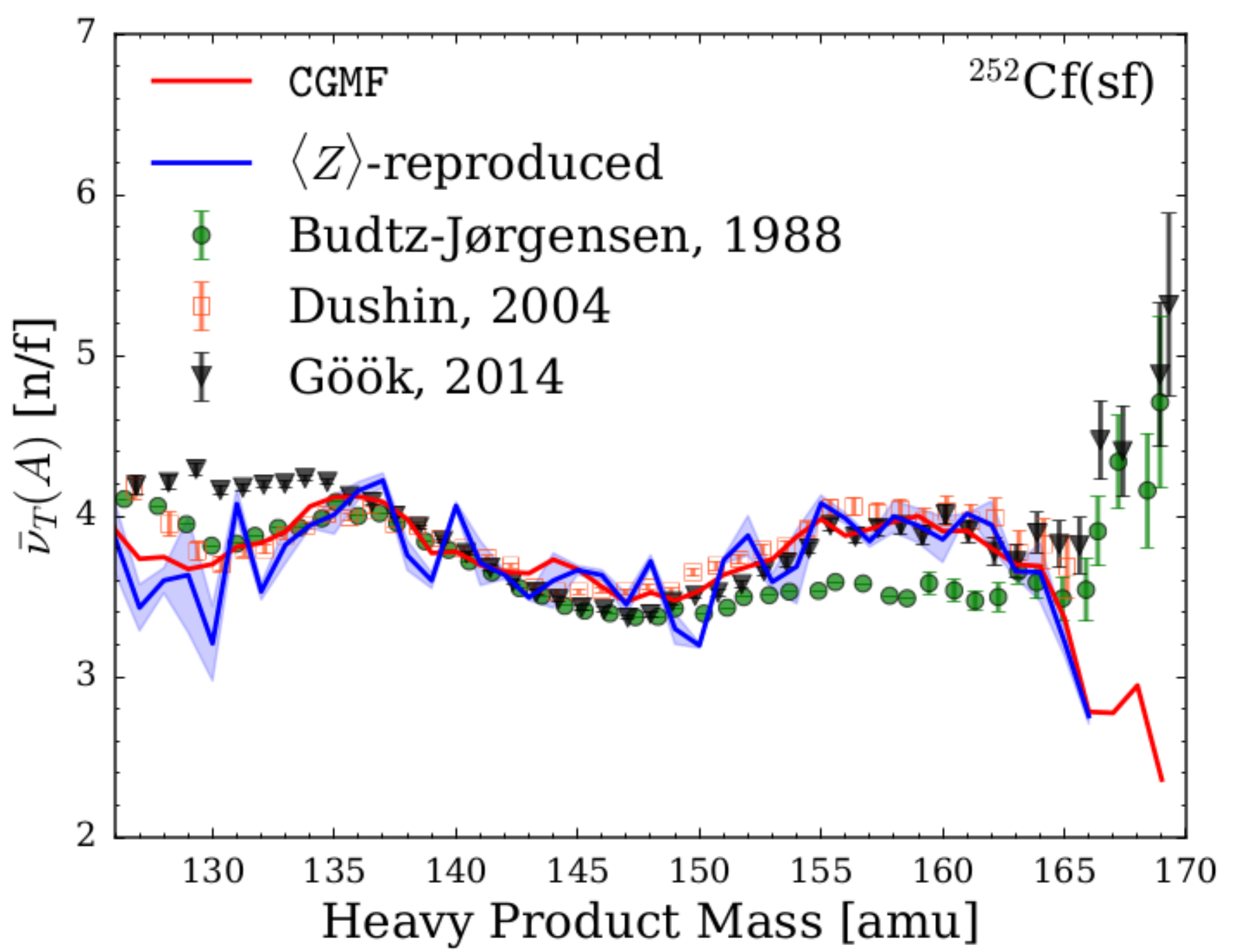}
\caption{\label{fig:AvgZProcess} Average total prompt neutrons emitted
  by a post-neutron heavy fission product with mass $A$ and its light complement in
  $^{252}$Cf(sf). The $\bar{\nu}_T(A)$ is shown for several
  experimental data sources ~\cite{budtz1988simultaneous,dushin2004facility,gook2014prompt} (points) and as calculated directly by the
  $\CGMF$ code (red). We use the IFY from $\CGMF$ to derive the
  $\bar{\nu}_T(A)$ through the average charge conservation (blue), which
  reproduces the direct $\bar{\nu}_T(A)$ with good agreement. The shaded regions indicate the errors on the
  derived $\bar{\nu}_T(A)$. Differences between the G{\"o}{\"o}k data
  and the $\CGMF$ results may indicate the need for charge-distributions
  near $A\sim 130$ different than Wahl or a different $\TKE(A)$. Similar
  results were generated with $^{235}$U(n$_\mathrm{th}$,f) and
  $^{239}$Pu(n$_\mathrm{th}$,f).
 }
\end{figure}

This verification has been conducted for the cases of $^{235}$U(n$_\mathrm{th}$,f),
$^{239}$Pu(n$_\mathrm{th}$,f), and $^{252}$Cf(sf), which all have extensive experimental
data on the various fission observables. The $\CGMF$ calculations require input data for
the mass, charge, average total kinetic energy $\TKE$, and spin $J$ distributions. The
$Y(A)$ and $\TKE (A)$ for $^{252}$Cf(sf) are from Ref.~\cite{gook2014prompt}, the $Y(Z)$
are chosen to follow the Wahl systematics~\cite{wahl2002systematics}, and the spin
distribution is a common parameterization~\cite{ohsawa1999multimodal,ohsawa2000multimodal}.
The remaining parameters, such as the spin-cutoff factor $\alpha$, are chosen to fit available
prompt neutron and $\gamma$-ray data. In the case of $^{252}$Cf(sf), the experimental
$\bar{\nu}$, $\bar{\nu}(A)$, and prompt neutron multiplicity distribution $P(\nu)$ are from
Refs.~\cite{mughabghab2006atlas,dushin2004facility,santi2008reevaluation}, respectively.
The $\CGMF$ calculation generates the IFY and we derive the $\bar{\nu}_T(A)$ via the outlined
procedure. This reconstructed $\bar{\nu}_T(A)$ and the
one generated directly from $\CGMF$ are shown for $^{252}$Cf(sf) in Fig.~\ref{fig:AvgZProcess}.

One can see that the $\langle Z\rangle$-reproduction of the $\bar{\nu}_T(A)$ is in good
agreement with the `true' values produced directly from the $\CGMF$ calculation. We
also note that our fit of the remaining parameters in $\CGMF$ to prompt neutron
observables results in good agreement with the experimental
data~\cite{budtz1988simultaneous,dushin2004facility,gook2014prompt}. The
experimental $\bar{\nu}(A)$ is often provided as a function of the primary fragment
mass $A^\prime$ (prior to neutron emission). We correct for this by mapping
$\bar{\nu}(A^\prime) \mapsto \bar{\nu}(A)$ with $A = A^\prime - \bar{\nu}(A^\prime)$
and interpolating over the product mass. The $\bar{\nu}_T(A)$ is then calculated numerically with
Eq.~\ref{eq:nubarT} under the constraint that $A + \bar{\nu}(A) + A_c + \bar{\nu}(A_c) = A_0$.
Using Eq.~\ref{eq:Nubar}, we have determined that the derived $\bar{\nu}$ is within $1\%$
of the value taken directly from $\CGMF$. Similar agreement was found for the
$^{235}$U(n$_\mathrm{th}$,f) and $^{239}$Pu(n$_\mathrm{th}$,f) reactions.

\section{\label{sec:NuCalc}Prompt Neutron Calculations}

\begin{figure}[h]
\centering
\includegraphics[width=\columnwidth]{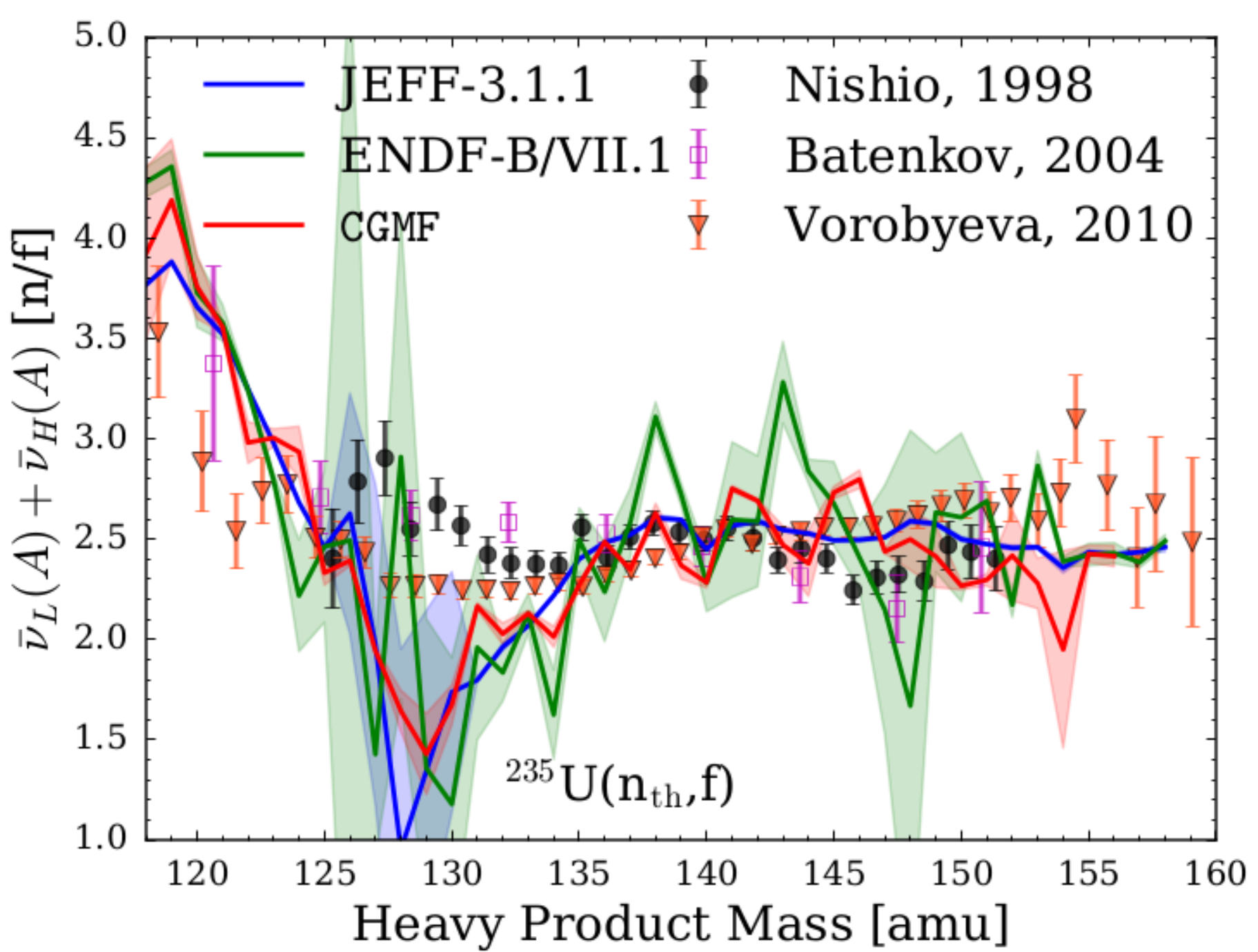}
\caption{\label{fig:U235NubarT} Average total prompt neutrons emitted
  by a post-neutron heavy fission product with mass $A$ and its light complement in
  the $^{235}$U(n$_\mathrm{th}$,f) reaction. The $\bar{\nu}_T(A)$ is
  shown for several experimental data sources~\cite{nishio1998multiplicity,Batenkov2005,vorobyev2009angular}
  (points) and as calculated by the average charge conservation method
  using the IFY from $\CGMF$ (red), ENDF-B/VII.1 (green), and JEFF-3.1.1
  (blue). The JENDL-4.0u2 evaluation is nearly identical to
  ENDF-B/VII.1. Shaded regions are the propagated errors on $\bar{\nu}_T(A)$.
  There is good agreement between the various calculations and data, but
  differences appearing near $A\sim 128$ could be due to differences in the
  charge-distributions.
 }
\end{figure}

We follow the procedure outlined above to determine both the $\bar{\nu}_T(A)$ and the
$\bar{\nu}$ values directly from the IFY of various fission reactions for the three major
libraries. In a few select cases, we were able to compare with experimental values. Figure~\ref{fig:U235NubarT}
shows the example of $^{235}$U(n$_\mathrm{th}$,f).
Here, we display the $\bar{\nu}_T(A)$ derived solely from the IFY of a $\CGMF$
calculation, the various evaluation libraries, and from experimental $\bar{\nu}(A)$ data~\cite{nishio1998multiplicity,Batenkov2005,vorobyev2009angular}. Again, we
see good agreement between the experimental data and our derivation of the
average total number of prompt neutrons emitted. We note that a similar degree of
consistency appears in the $^{233}$U(n$_\mathrm{th}$,f) and
$^{239}$Pu(n$_\mathrm{th}$,f) reactions when comparing with their respective
$\bar{\nu}(A)$ data from Ref.~\cite{Nishio1995,tsuchiya2000simultaneous,Batenkov2005}
and Ref.~\cite{apalin1965neutron,fraser1966nuclear,nishio1998multiplicity}.

We can calculate $\bar{\nu}$ via Eq.~\ref{eq:Nubar} using the $\langle Z\rangle$-derived
$\bar{\nu}_T(A)$ for the evaluated yields. We have verified that there is a negligible difference
when summing over the light, heavy, or averaging over both in Eq.~\ref{eq:Nubar}. Listed in Tab.~\ref{tab:nubar},
for several fission reactions, are the calculated $\bar{\nu}$ for the  three major evaluation
libraries and experimental data for comparison. Overall, one can see that only about half of the listed reactions show good
agreement with the experimental values in at least one of the evaluation
libraries. This result shows a fundamental inconsistency in many of the
IFY with the $\bar{\nu}$ data.
\begin{table*}[t]
\centering
\renewcommand{\arraystretch}{1.25}
    \begin{tabular}{|c||c|c|c|c|c|c|}
        \hline
	 & $^{233}$U(n$_\mathrm{th}$,f) &$^{235}$U(n$_\mathrm{th}$,f) & $^{238}$U(n$_\mathrm{f}$,f) & $^{239}$Pu(n$_\mathrm{th}$,f) & $^{241}$Pu(n$_\mathrm{th}$,f) & $^{252}$Cf(sf) \\ \hline \hline
	ENDF-B/VII.1 & $\boldsymbol{2.30\pm 0.09}$ & $2.43\pm 0.09$ & $\boldsymbol{4.29\pm 0.13}$ & $2.83\pm 0.08$ & $2.97\pm 0.09$ & $3.67\pm 0.11$ \\ \hline
	JENDL-4.0u2 & $\boldsymbol{2.30\pm 0.09}$ & $2.43\pm 0.09$ & $\boldsymbol{4.27\pm 0.14}$ & $2.82\pm 0.08$ & $2.97\pm 0.10$ & $3.66\pm 0.12$ \\ \hline
	JEFF-3.1.1 & $\boldsymbol{2.40\pm 0.05}$ & $2.40\pm 0.05$ & $4.72\pm 0.09$ & $\boldsymbol{2.97\pm 0.05}$ & $\boldsymbol{3.10\pm 0.07}$ & $\boldsymbol{4.64\pm 0.09}$ \\ \hline
	Exp. & $2.49\pm 0.004$ & $2.43\pm 0.003$ & $4.69\pm 0.159$~\cite{laurent2014new} & $2.88\pm 0.006$ & $2.92\pm 0.007$ & $3.77\pm 0.004$ \\ \hline
    \end{tabular}
\vspace{3mm}
\renewcommand{\arraystretch}{1.25}
    \begin{tabular}{|c||c|c|c|c|c|c|c|}
        \hline
	 & $^{237}$Np(n$_\mathrm{th}$,f) & $^{238}$Np(n$_\mathrm{i}$,f) & $^{238}$Pu(n$_\mathrm{i}$,f) & $^{241}$Am(n$_\mathrm{th}$,f) & $^{243}$Cm(n$_\mathrm{th}$,f) & $^{244}$Cm(n$_\mathrm{i}$,f) & $^{245}$Cm(n$_\mathrm{th}$,f) \\ \hline \hline
	ENDF-B/VII.1 & $1.94\pm 0.07$ & $2.37\pm 0.09$ & $1.95\pm 0.07$ & $2.71\pm 0.10$ & $2.77\pm 0.10$ & $3.63\pm 0.13$ & $3.13\pm 0.10$ \\ \hline
	JENDL-4.0u2 & $1.93\pm 0.07$ & $2.37\pm 0.09$ & $1.93\pm 0.07$ & $2.82\pm 0.08$ & $2.74\pm 0.10$ & $3.61\pm 0.13$ & $3.11\pm 0.11$ \\ \hline
	JEFF-3.1.1 & $2.74\pm 0.06$ & $2.87\pm 0.09$ & $3.09\pm 0.10$ & $3.35\pm 0.08$ & $3.76\pm 0.09$ & $4.04\pm 0.12$ & $4.30\pm 0.09$ \\ \hline
	Exp. & $2.52\pm 0.016$ & $2.77\pm 0.14$~\cite{howerton:1977} & $2.88\pm 0.14$~\cite{howerton:1977} & $3.21\pm 0.032$ & $3.43\pm 0.047$ & $3.33\pm 0.17$~\cite{howerton:1977} & $3.72\pm 0.06$ \\ \hline
    \end{tabular}
\caption{\label{tab:nubar} List of derived $\bar{\nu}$ for various fission reactions
  from ENDF/B-VII.1, JENDL-4.0u2, and JEFF-3.1.1. Considered neutron
  reactions are at thermal ($\mathrm{n_\mathrm{th}}$), intermediate
  ($E_\mathrm{n_i} \approx 500\,\mathrm{keV}$), or fast ($E_\mathrm{n_f}=14\,\mathrm{MeV}$) energies. Experimental $\bar{\nu}$ are from
  Ref.~\cite{mughabghab2006atlas}, unless otherwise noted. The
  `experimental' $\bar{\nu}$ for $^{238}$Np(n$_\mathrm{i}$,f),
  $^{238}$Pu(n$_\mathrm{i}$,f), and $^{244}$Cm(n$_\mathrm{i}$,f) were
  calculated via Howerton's parameterization~\cite{howerton:1977} using
  the binding energies derived from Ref.~\cite{Moller:2015fba}. Major
  fission reactions (top) showed moderate agreement between experimental
  data and those derived from the evaluated yields, but some values (bolded)
  fall outside the $1\sigma$ errors. Less known or studied fission reactions
  (bottom) seldom showed agreement. Errors on the derived $\bar{\nu}$
  are propagated from the fission yields.}
\end{table*}

Changes in $\langle Z\rangle(A)$ have a dramatic effect on the derived
$\bar{\nu}_T(A)$. For example, if we artificially increase the $\langle Z\rangle(A)$
by $0.28\%$ $(0.65\%)$ for $A_h \in [129,134]$ in the JEFF-3.1.1 (ENDF-B/VII.1) evaluations of $^{233}$U(n$_\mathrm{th}$,f),
then $\bar{\nu}$ becomes $\sim 2.49$ for each library, consistent with Ref.~\cite{mughabghab2006atlas},
and the $\bar{\nu}_T(A)$ in this mass region match Ref.~\cite{apalin1965neutron,fraser1966nuclear,nishio1998multi} better.
Identifying the exact source of a disagreement between
the experimental and derived $\bar{\nu}$ is difficult, as a large range of product
masses contribute. Even so, discontinuities in $\langle Z\rangle(A)$ can identify
charge distributions that display anomalous behavior, such as the extremely low
yield of $^{128}$Te in the JENDL-4.0u2 and ENDF-B/VII.1 evaluations of
$^{233}$U(n$_\mathrm{th}$,f).

\begin{figure}[h]
\centering
\includegraphics[width=\columnwidth]{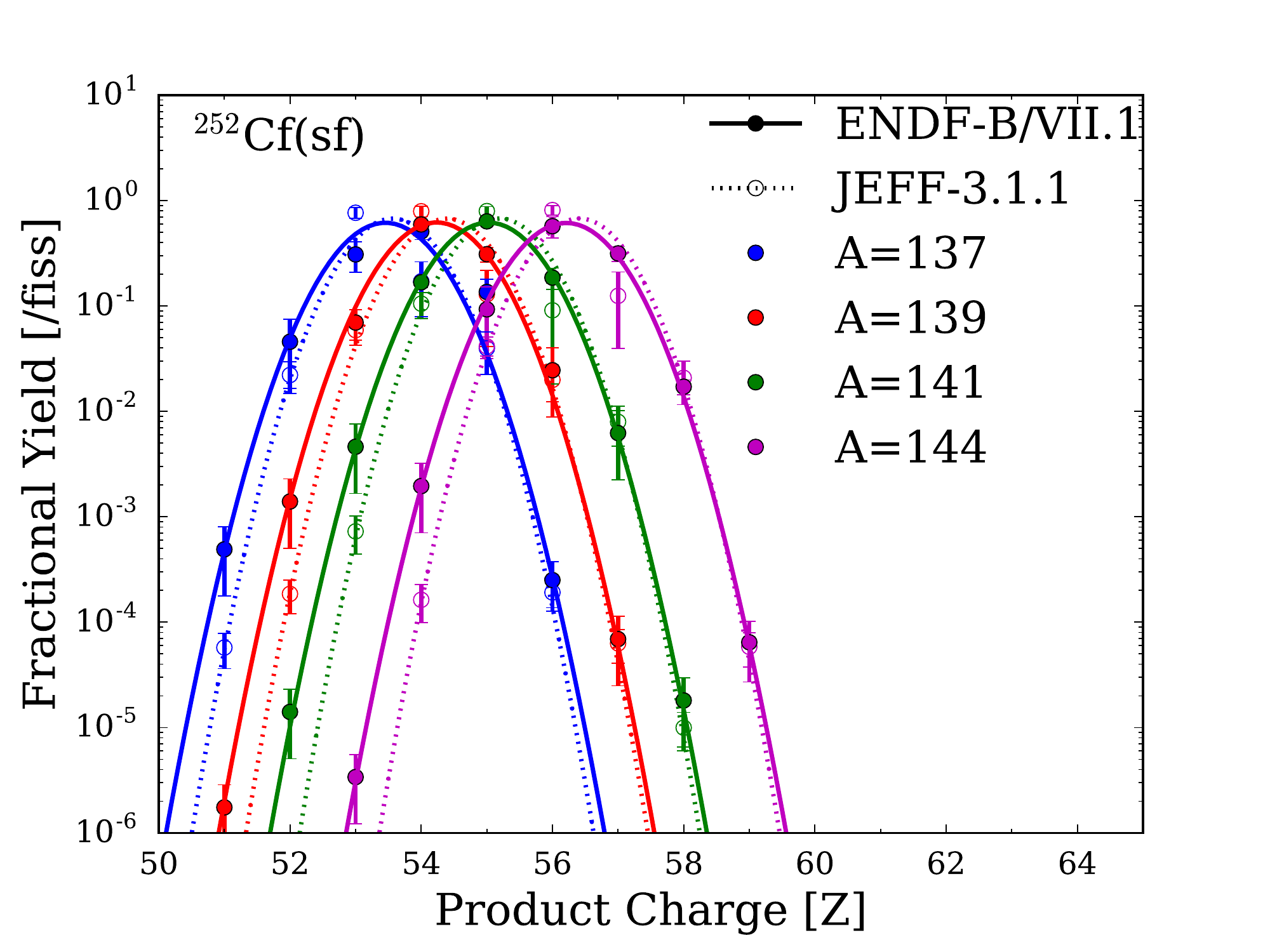}
\caption{\label{fig:AvgZShift} Fitted charge distributions with a Gaussian
  representation for a few heavy product masses in the ENDF-B/VII.1
  (solid) and JEFF-3.1.1 (dotted) evaluations of $^{252}$Cf(sf). The closed and
  open points represent the evaluated ENDF-B/VII.1 and JEFF-3.1.1
  data, respectively. One can observe a systematic shift to larger
  $\langle Z\rangle$ in JEFF-3.1.1 as compared with ENDF-B/VII.1.
 }
\end{figure}

A noticeable trend from Tab.~\ref{tab:nubar} is that the JEFF-3.1.1 evaluations
tend to generate larger values of $\bar{\nu}$ than JENDL-4.0u2 and ENDF-B/VII.1,
such as in the $^{252}$Cf(sf) calculations, where a difference of almost an
entire neutron exists. As
mentioned before, the scaling of the mean charge has a large impact on the
derived $\bar{\nu}$. An increase in $\langle Z\rangle(A)$ will require a lower
$\langle Z\rangle(A_c)$. As the function $\langle Z\rangle(A)$ is nearly linear,
this will lead to a lower $A_c$ and larger $\bar{\nu}$. Thus, a fission yield
evaluation with a larger $\langle Z\rangle(A)$, such as the trend seen in
Fig.~\ref{fig:AvgZShift}, will generate a larger $\bar{\nu}$. The $\langle Z\rangle(A)$ for $^{252}$Cf(sf) is, on average, $\sim 0.4\%$ larger
in JEFF-3.1.1 than ENDF-B/VII.1. While this change may seem small, the
difference is summed over all masses. When we correct for this
effect, $\bar{\nu}$ becomes consistent among the three libraries. Experimental
re-measurements of suspect yields should be able to identify the correct charge
distribution, as the differences between libraries discussed here are at the
percent-level.
Thus, the accurate measurements of $\bar{\nu}$, along with
accurate $\TKE$ data, could help to identify issues in the evaluated IFY, such
as the increased mean charge function $\langle Z\rangle(A)$ in the JEFF-3.1.1
evaluation of $^{252}$Cf(sf). We also note a similar increase in $\langle Z\rangle(A)$
for JEFF-3.1.1 relative to ENDF-B/VII.1 in the $^{237}$Np(n$_\mathrm{th}$,f),
$^{238}$Pu(n$_\mathrm{i}$,f), $^{241}$Am(n$_\mathrm{th}$,f),
$^{243}$Cm(n$_\mathrm{th}$,f), $^{244}$Cm(n$_\mathrm{i}$,f), and
$^{245}$Cm(n$_\mathrm{th}$,f) reactions, which also show
larger derived $\bar{\nu}$. This result indicates that yields evaluations utilizing
average charge parameterizations~\cite{wahl2002systematics} may need to
take the resulting $\bar{\nu}$ into consideration and lower or raise
$\langle Z\rangle(A)$ accordingly.

Several fission reactions have evaluations over a range of incident neutron
energies. Examining the $\langle Z\rangle(A)$ for a single fission reaction, we see
an increase in the mean charge as the incident neutron energy increases, which
causes an increase in $\bar{\nu}$ as expected. The JEFF-3.1.1 library
contains a thermal, intermediate, and fast yields evaluation for $^{235}$U(n,f).
Using our method, we derive $\bar{\nu}=[2.40\pm 0.05$, $2.46\pm 0.04$, $4.33\pm 0.08]$
for the three incident energies, which are in agreement with
$\bar{\nu} = [2.407\pm 0.0066, 2.461\pm 0.0028, 4.387\pm 0.0081]$~\cite{Manero:1972}.

\section{Conclusion}
We have conducted a detailed survey of various fission
reactions at several incident neutron energies and across three major
yield evaluations: ENDF-B/VII.1~\cite{chadwick2011endf}, JENDL-4.0u2~\cite{shibata2011jendl},
and JEFF-3.1.1~\cite{koning2006jeff}. From our initial analysis,
we find that all yield evaluations satisfy basic charge and mass
conservation. A handful of nuclides showed strong disagreement
with the Wahl charge distributions~\cite{wahl2002systematics}:
$^{86}$Ge, $^{88}$As, $^{92,100,105}$Zr, $^{134}$Te, and $^{136}$Xe in
the $^{235}$U(n$_\mathrm{th}$,f) reaction (mostly in ENDF-B/VII.1),
$^{103}$Nb, $^{106}$Mo, $^{108}$Tc, $^{110}$Ru, $^{137}$I, $^{139}$Xe,
$^{141}$Cs, and $^{144}$Ba in the $^{252}$Cf(sf) reaction (mostly in
JEFF-3.1.1), and $^{98}$Zr, $^{136}$Xe, and $^{142}$Ba in the
$^{233}$U(n$_\mathrm{th}$,f) reaction (all libraries). A few
instances of charge asymmetry were identified, but could be remedied
by adjusting anomalous yields in the symmetric regions
to fit a Gaussian charge distributions.

We derived a simple average charge conservation method to infer the prompt
neutron multiplicity from independent fission product yields. This method
was shown to be accurate by using the fully self-consistent $\CGMF$
code, which reconstructed $\bar{\nu}$ within $1\%$ of the `true'
value. This method was applied to the suite of fission reactions and
incident neutron energies supported by ENDF-B/VII.1, JENDL-4.0u2,
and JEFF-3.1.1. We find that the major fission reactions could produce
$\bar{\nu}(A)$ distributions and $\bar{\nu}$ values in reasonable agreement with
experimental data. For less known or studied fission reactions, very few
yield evaluations were consistent with prompt neutron data. We have
identified possible sources of these discrepancies, such as the anomalous
increase in $\langle Z\rangle(A)$ for the JEFF-3.1.1 $^{252}$Cf(sf) evaluation,
relative to ENDF-B/VII.1 or JENDL-4.0u2, which dramatically affects the
derived $\bar{\nu}$. Specific nuclides have been identified as having
anomalous yields and could be remeasured in future experiments~\cite{Meierbachtol201559,refId0}.
New experiments looking to measure $A$, $Z$, $\mathrm{TKE}$, and the
prompt neutron emission simultaneously~\cite{Martin2015} could also
provide experimental validation of the consistency checks presented here.
Finally, new evaluated fission product yields may be necessary, a goal
already identified by the nuclear data community~\cite{koning2009nuclear}
and demonstrated in Ref.~\cite{kawano2013estimation}. These
self-consistency checks and a more rigorous uncertainty
quantification~\cite{DACRUZ2014531} should be an integral part of the new
evaluated yields.

\acknowledgements
The author would like to thank Patrick Talou, Ionel Stetcu, and Alejandro
Sonzogni for helpful discussions and comments.
This work was supported by Los Alamos National Security LLC,
U.S. National Nuclear Security Administration, and 
U.S. Department of Energy. This publication is LA-UR-17-27790.



\let\clearpage\relax


\end{document}